# From Pragmatic to Systematic Software Process Improvement: An Evaluated Approach

Marco Kuhrmann[1,2], Daniel Méndez Fernández[2]

[1]University of Southern Denmark, The Mærsk Mc-Kinney Møller Institute, 5230 Odense, Denmark
   E-Mail: kuhrmann@mmmi.sdu.dk

[2]Technische Universität München, Faculty of Informatics, 85748 Garching, Germany
   E-Mail: mendezfe@in.tum.de

Correspondence to: Marco Kuhrmann, kuhrmann@mmmi.sdu.dk

## Abstract

Software processes improvement (SPI) is a challenging task, as many different stakeholders, project settings, and contexts and goals need to be considered. SPI projects are often operated in a complex and volatile environment and, thus, require a sound management that is resource-intensive requiring many stakeholders to contribute to the process assessment, analysis, design, realisation, and deployment. Although there exist many valuable SPI approaches, none address the needs of both process engineers and project managers. This article presents an Artefact-based Software Process Improvement & Management approach (ArSPI) that closes this gap. ArSPI was developed and tested across several SPI projects in large organisations in Germany and Eastern Europe. The approach further encompasses a template for initiating, performing, and managing SPI projects by defining a set of 5 key artefacts and 24 support artefacts. We present ArSPI and discus results of its validation indicating ArSPI to be a helpful instrument to set up and steer SPI projects.

## 1 Introduction

Software processes comprise many process assets, which need to be designed, implemented, quality assured, and managed in the context of an organisation-wide software process management (SPM; [27]). Software process improvement (SPI; [12]) aims at the systematic analysis, re-/design, and evaluation of a particular process. As part of SPM, it forms an important step for organisations of all sizes to succeed in the market [11]. However, SPI is costly and improved processes need time to be disseminated, making the impact of SPI hard to measure and justify [4], [6], [8]. Therefore, and because of the associated costs, many software managers are reluctant to conduct SPI [6], or companies give up SPI at all [10]. Niazi et al. [28] mention the importance of an effective strategy to successfully implement SPI, what is especially true for the management of a process's evolution after its initial deployment, i.e. changes must be tracked [29] and the evolution of external standards must be considered in the process maintenance [23]. Furthermore, a sound project organisation, i.e. allocating resources, defining deliverables, or tracking progress, is crucial to set up a systematic SPI [3].

From the perspective of a process engineer, we still miss a guidance to conduct a flexible but systematic SPI project going beyond purely assessment-driven approaches, like CMMI [5] or ISO/IEC 15504 [14]. That is, we need support for process engineers to systematically organise and perform SPI projects in the context of an SPM strategy, while leaving open the way of conducting particular improvements.

In this article, we contribute a model for an *Artefact-based Software Process Improvement & Management* (ArSPI). The model emerges from initial pragmatically conducted improvement activities [18]. Based on our experiences across several SPI projects in large organisations in Germany, we inferred and systematised our approach, which we applied and validated in practice in Germany and Eastern Europe. Our approach relies on the principle of artefact orientation [25]. That is, by concentrating on the key artefacts to be created in SPI projects, we abstract from actual SPI activities. We thus give process engineers the freedom to apply methods appropriate for their particular situation while being able to clearly define the interfaces to supporting activities, e.g., quality management. ArSPI defines a template that process engineers can use to set up and manage SPI projects.

The remainder of this article is organised as follows. In Section 2, we discuss related work, which gaps are left open, and how we intent to close those gaps. In Section 3, we introduce our approach in detail, before giving a concluding summary in Section 4.

## 2 Related Work

In this article, we present an SPI method that, compared to existing models, follows an alternative approach. Instead of focussing on assessments or specific improvement procedures, our model is based on the principle of artefact orientation [25]. According to Frailey [9], relying on artefacts is advantageous as artefacts ease, inter alia, the creation of a common terminology. In a study on the perception of artefact-based software processes [20] and in an experiment on the perception of process modelling [21], we further found indicators supporting Frailey's conclusion. Reviewing available and well-disseminated SPI models, e.g., CMMI [5], ISO/IEC 15504 [14], and ISO/IEC 12207 [13], we find, however, that the focus in current approaches lies on providing comprehensive descriptions of principles and procedures rather than on providing precise artefact models. Furthermore, these models are often criticised to be too voluminous, too complex, or to result in processes that might lead to an improvement alien to the organisation [2], [26], [32]. In response to this shortcoming, tailored variants of these models aim at better addressing small and very small companies, e.g., ISO/IEC 29110 [15]. Those approaches, however, remain normative and they usually focus on process assessments only. Even the recently published standard ISO/IEC 33014 [16] focuses on activities without precisely defining the required artefacts. Apart from the standards, several method proposals were made that rely on best practices and standards emphasising needs of small companies. PROCESSUS [11], BG-SPI [1], LAPPI [31], and BOOTSTRAP [24], are representatives of such methods. These proposals comprise activity-based guidelines providing detailed procedures process engineers should follow (e.g., [11], [1], and [31]), or they aim to simplify process assessments (e.g., [24]). Artefacts are mentioned (e.g., PROCESSUS and BG-SPI), but detailed models of artefact structures and relationships are not provided. Activity-based approaches therefore encounter problems in practice: *What if the described order of activities does not meet the needs of the actual context?* A missing description of the expected outcomes hampers learning curves of process engineers, limits the comparability of SPI projects, and thereby limits the opportunities to create reusable assets for enhancing improvement processes, which are all aspects investigated in the context of SPI success factors (e.g., Melzer and Stellis [33]) and human aspects in SPI (e.g., Viana et al. [34]).

Our proposed approach precisely defines the artefacts allowing for creating a model of the expected results that can be tested, e.g., for completeness and consistency. Furthermore, an artefact-based approach allows for bridging the gap between single SPI project instances and organisation-wide SPI programs to provide SPI projects with a stable environment as, for instance, recommended by Rainer and Hall [30]. Artefact models remain stable and only the respective methods for the artefact creation need to be adapted for the respective context, which allows for, e.g., a flexible tailoring of improvement endeavours as considered crucial by Melzer and Stellis [33]. The subsequently presented ArSPI[1] model provides scalable and adaptable SPI project- and artefact templates [17] supporting process engineers to set up and organise SPI projects. ArSPI defines a method-agnostic, but general structure (the embodiment with particular methods is out of scope) that process engineers can use in combination with their preferred methods as, e.g., previously contributed for the RE improvement domain [26]. This article is based on previously published material: We first presented ArSPI as method proposal [19], gave a brief overview of the key concepts, and provided two practical examples. The full ArSPI model, i.e. all UML models, tailoring profiles, and so forth, are documented in our complementing technical report [17], and the overall construction procedure of ArSPI can be depicted from [18]. In this article, we extend the presentation of ArSPI by providing more details and background, and we provide information on the validation of ArSPI in academia and practice.

---

[1] The ArSPI website: http://www4.in.tum.de/~kuhrmann/arspi.shtml

# 3    A Model for Artefact-based Software Process Improvement & Management

We introduce ArSPI by describing the artefact model in Section 3.1, and the life cycle model in Section 3.2. We concentrate on an overview and present selected concepts. Further details can be depicted from [17]. An overview of the overall evaluation strategy and a discussion on selected results is given in Section 3.3.

The ArSPI model is an artefact-based approach to organise SPI projects. Its nature puts emphasis on the artefacts being produced and used in SPI projects – it thereby focuses on the "what" rather than dictating which methods to apply in which sequence. Hence, ArSPI defines a comprehensive, but flexible model that addresses SPI projects as well as organisation-wide SPM.

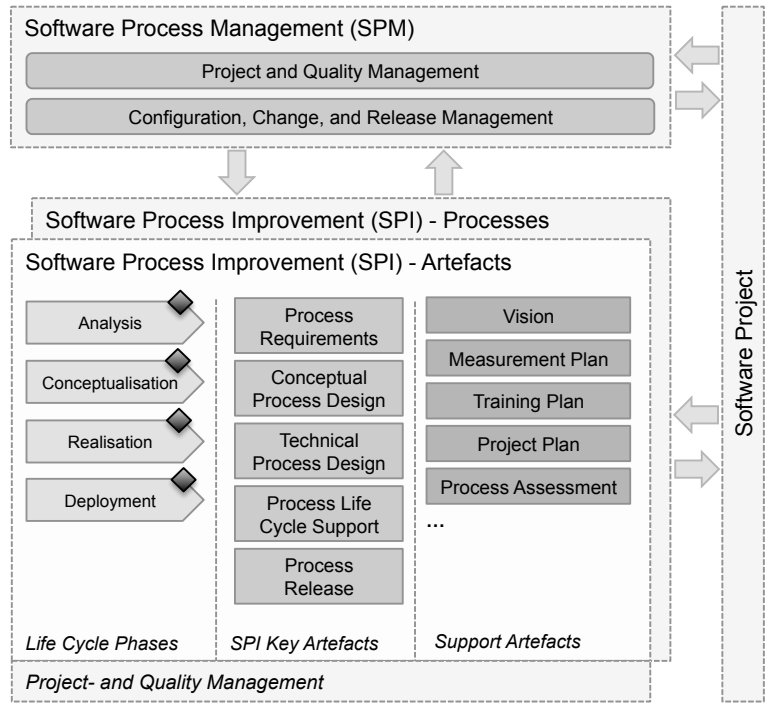

**Figure 1 Overview of the ArSPI model.**

ArSPI consists of three parts (Figure 1):

- **SPI Projects:** ArSPI characterises SPI projects by defining 5 mandatory *key artefacts* (Table 1), 24 complementing *support artefacts* (Table 2), *life cycle phases* to which the artefacts are assigned for the project organisation, and a *process model*, which remains rudimentary due to the artefact-based nature of ArSPI.
- **Organisation-wide SPM:** An organisation-wide SPM implements the SPI strategy, owns the software processes, initiates particular improvement endeavours, and deploys process releases. For this, ArSPI defines artefacts and processes to (1) define interfaces between the organisation-wide SPM and particular SPI projects, and (2) to establish the necessary management and administration processes.
- **Software Projects:** SPI primarily targets software projects where (improved) processes are applied and that serve as major source to gather experiences and the data necessary to conduct further improvements.

The ArSPI model provides a framework, which supports companies in implementing SPM including organisation-wide improvement programs, particular improvement projects, and the management of processes. By defining interfaces and fine-grained artefact structures, ArSPI bounds the different parts of SPM together and provides a unified perspective.

## 3.1    The ArSPI Artefact Model

We describe the artefact model of ArSPI in more detail by defining the key artefacts. Furthermore, we provide the essence of how the artefacts are designed and how particular artefacts materialise in projects.

As described in [18], ArSPI defines 5 key artefacts (Table 1), which have to be created in every SPI project. 24 support artefacts, created in response to particular project requirements, e.g., assessment and quality assurance artefacts, accompany these key artefacts. Table 2 lists selected support artefacts and provides a brief description. The complete list of support artefacts can the depicted from [17].

**Table 1 Overview of the ArSPI key artefacts.**

| ID | Artefact Description | SPI Project Life Cycle Phase |
|---|---|---|
| PRQ | The *Process Requirements* artefact contains all requirements regarding the process. To collect all relevant requirements, the PRQ defines the following top-level structure:<br>• Goals<br>• Stakeholders and Roles<br>• Requirements<br>• Overall Process Draft<br>• Technical Infrastructure<br>• Basic Conditions | Analysis |
| CPD | The *Conceptual Process Design* contains all designs of a process without paying attention to any technical realisation. It refines all process-related requirements and transfers them into concrete processes and process parts. It defines the following top-level structure:<br>• Goals (shared with PRQ)<br>• Principles<br>• *Planned Adaptations*: Organisation and Roles, Artefacts, and Processes<br>• *Additional Requirements*: Tailoring, Process Documentation, and Supporting Material<br>• Requirements Tracing (shared with PRQ and TPD) | Conceptualisation |
| TPD | The *Technical Process Design* refines the CPD regarding a concrete technical realisation and tool/tool infrastructure to be used for its realisation:<br>• (Refinement of the CPD structure, cf. Figure 2)<br>• Logical and Physical Model Organisation | Realisation |
| PLC | The *Process Life Cycle Support* comprehends all information, agreements, and definition regarding complementing processes supporting the deployment, training, and further development of a concrete process as well as its evaluation and measurement:<br>• Training<br>• Deployment and Further Development<br>• Measurement and Evaluation<br>• Change Management | *Created early in the Analysis phase, at latest in the Deployment phase* |
| PR | A *Process Release* is a concrete process package that is shipped and deployed. The results produced in the SPI project dynamically define the PR's structure. | Deployment |

**Table 2 Selection of the ArSPI support artefacts.**

| Artefact | Description |
|---|---|
| User Evaluation Plan | While the measurement plan of a process aims at measuring the process performance in general, a *User Evaluation Plan* aims to evaluate the actual use of a process. In contrast to "classic" KPI-based measuring, the user evaluation is more of a qualitative nature. |
| Training Material | *Training Material* consists of material to train the process consumers. The *Training Material* is specific for certain user groups/stakeholders and for particular *Process Releases*. Usually, *Training Material* is explicitly defined for stakeholder groups and, therefore, provides different perspectives and information at different levels of abstraction. |
| SPL-Delta Report | If the considered process is based on a software process line (SPL), a delta report is helpful to analyse deviations from the SPL base process to support long-term development (having the SPL's evolution in mind), and to support compliance assessments. |

ArSPI's artefact model defines the structure of the particular artefacts. A comprehensive set of associations connects the artefacts with each other to allow for refinements and tracing, e.g., which requirements are how designed and realised. Figure 2 shows an example of the UML model in which the *Conceptual Process Design* and *Technical Process Design*, and their relationships are illustrated.

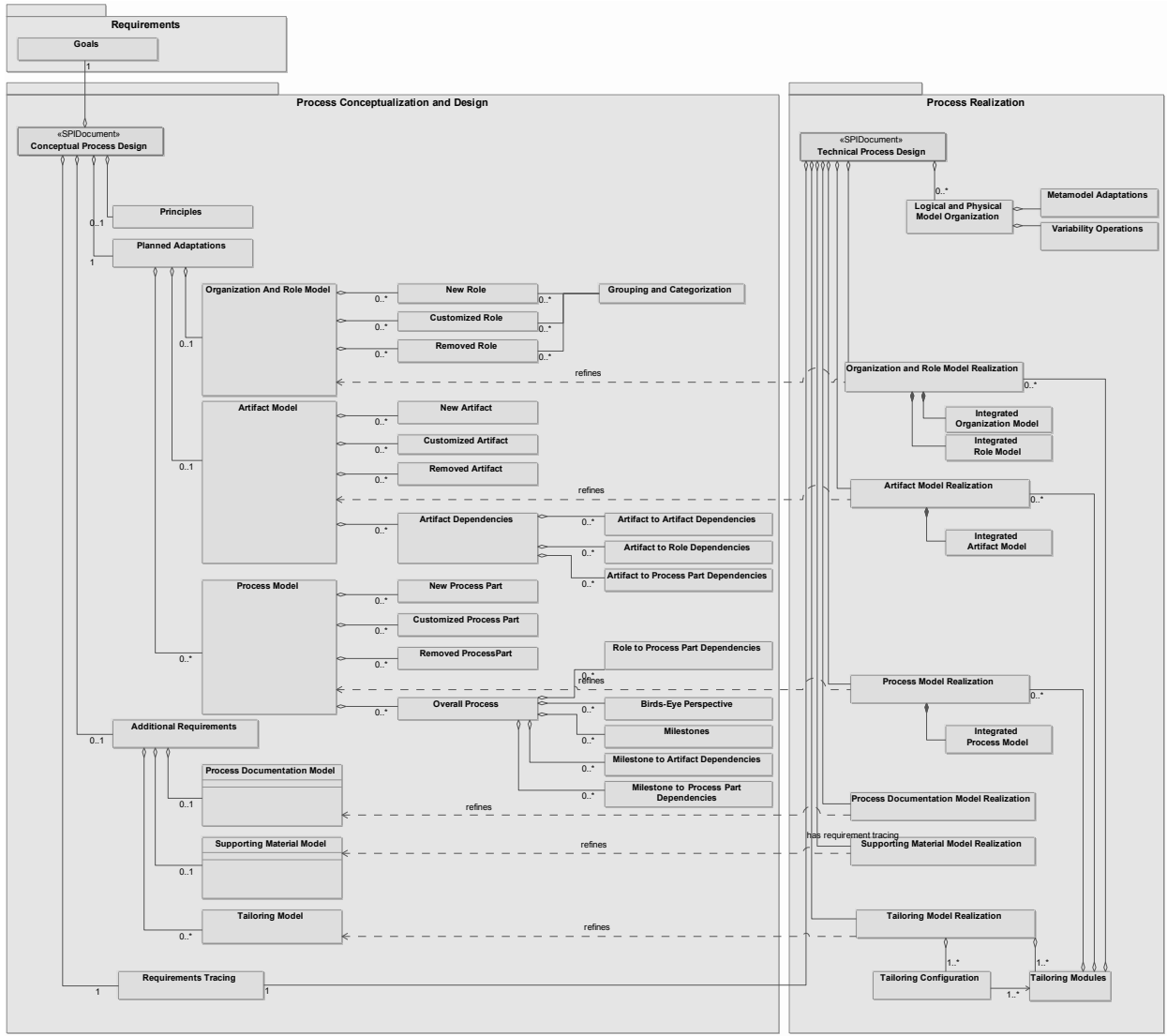

**Figure 2 Excerpt of the ArSPI model: the CPD and TPD artefacts, and their associations.**

The UML model comprehends all elements identified during the construction procedure of ArSPI [18], and integrates them into a harmonised and consistent model. Hence, the model allows for setting up basic quality assurance measures, e.g., completeness, and, by instrumenting the relationships, for checking consistency of the results. Furthermore, we decided to use UML as modelling language to serve several realisation options. In the simple case, the artefact aggregation structure can be easily transformed into a document structure, e.g., a Word template, which process engineers can use to document process requirements. Likewise, the artefact model can also be instrumented in tools (e.g., for design and enactment). Depending on the actual context and the used (project) infrastructure, ArSPI artefacts can thereby materialise as documents or computable data of corresponding tools. Therefore, ArSPI provides a structured collection of concepts that can be tailored according to the respective context.

## 3.2 The ArSPI Life Cycle Model

We briefly describe the life cycle model of SPI projects and the overall organisation model, show how ArSPI integrates SPI and SPM, and we provide insights into the implementation of ArSPI.

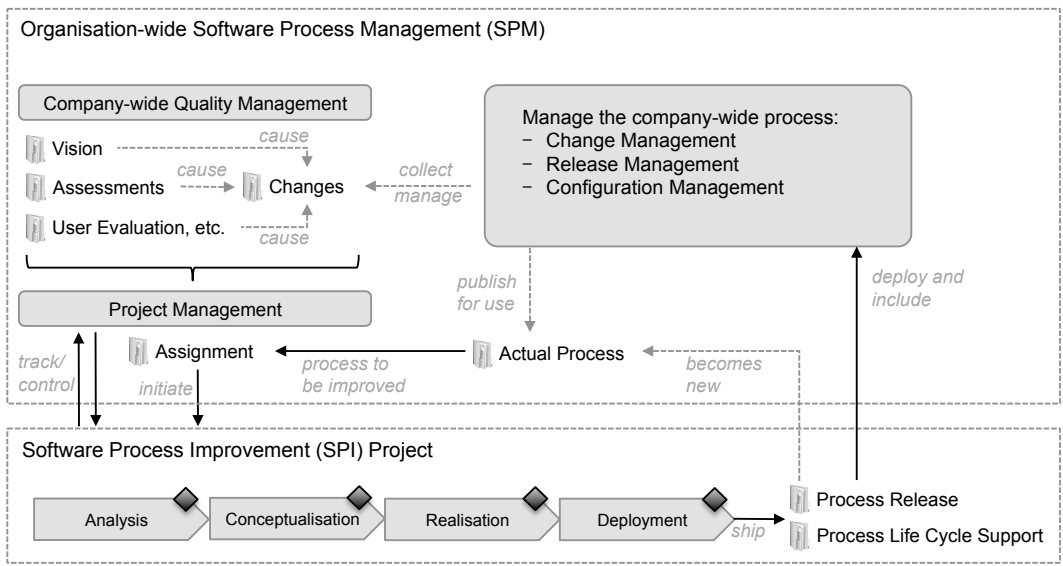

**Figure 3 The ArSPI life cycle and organisation model (simplified view).**

Figure 3 presents a simplified view of the life cycle model, which binds together an organisation-wide SPM and specific SPI projects. The main life cycle is visualised by the solid lines, examples of internal information flow are visualised by the dotted lines. SPI projects are always embedded into an organisational context, which provides a *Vision* as part of the overall process-related strategy, and handles configuration-, release-, and change management of a process that is subject to continuous improvement. Furthermore, the organisation initiates SPI projects.

### 3.2.1 Organisation Point of View

An SPI project starts with a *Project Assignment* (e.g., a contract) from the process-owning organisation, and is iteratively performed by the process engineers. Based on the Plan-Do-Check-Act cycle [7], iterations comprise up to four phases. The goal is to deploy one *Process Release* per iteration. *Process Releases* and *Process Life Cycle Support* documentation are shipped to the organisation that includes a release in the *release management* (combined with a *configuration management*), and, eventually, publishes a release as new *Actual Process* for use in projects. A *change management* is enabled for the new *Process Release*, and, together with a *quality management*, collects required *Changes* for further improvement cycles. The *Process Life Cycle Support* artefact comprises all procedures necessary to establish necessary management tasks (Table 1). The *quality management* also manages the *Vision* representing the overall improvement goals, e.g., a certain CMMI level. A *Vision*, a set of *Changes*, and an *Actual Process* as reference are the basic inputs to initiate improvement cycles.

**Example.** This example is extracted from a long-term industry project (Table 3, study 5). The example boils down more then 5 years of cooperation in which the organisation-wide standard software process of a German government agency was improved in several iterations. The agency adopted the V-Modell XT to implement the contracting and development processes, i.e. the process is part of the V-Modell XT process line [23]. The agency's process variant is derived from the "V-Modell XT Bund", which itself is a variant of the generic "V-Modell XT Reference Model". Figure 4 illustrates how the different parties interface, and how organisation-internal and external triggers affecting SPI projects demand for a mature process management.

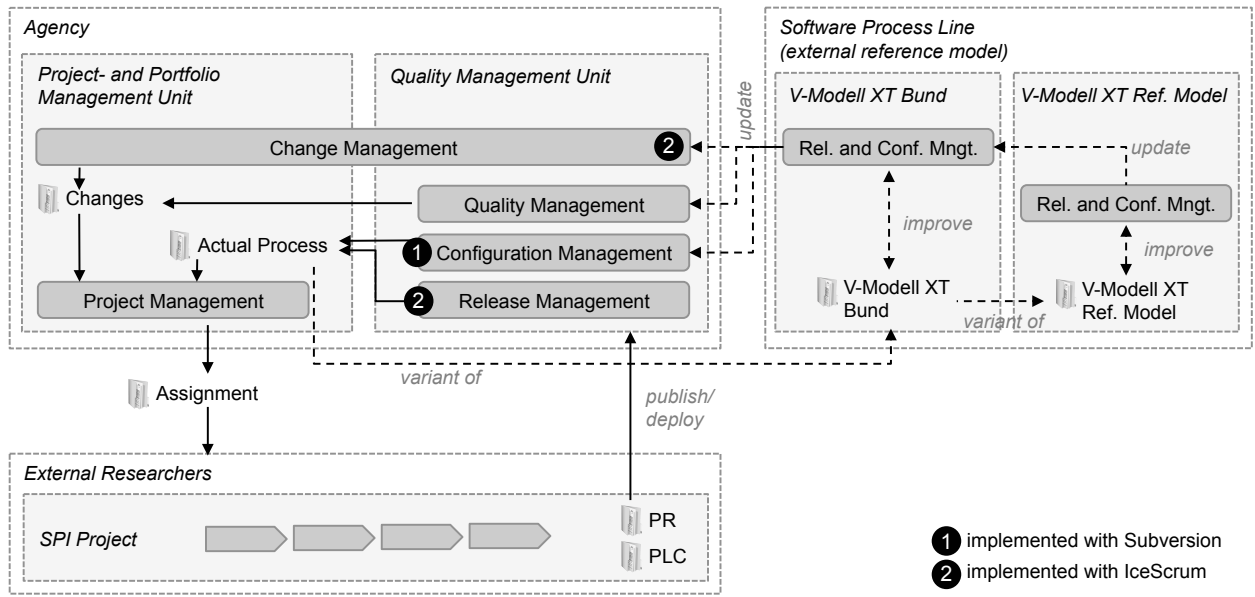

**Figure 4 ArSPI in an organisation-wide SPM.**

The agency (Figure 4, left side) has its own feedback and improvement cycles. Hence, it internally triggers the process's evolution. Software developers and project managers report problems or propose improvements, which are managed in an IceScrum[2] system. The portfolio management and quality management units owning the process bundle change requests and (new) requirements to direct another iteration in the improvement program. Finally, an SPI project is initiated (cf. Section 3.2.2). The SPI projects each generate several review versions, an internal beta version, a public release candidate, and eventually the *Process Release* (PR). Currently, this agency deploys one major PR per year.

While the agency is in full control of its own process variant and directs the improvement, the variant as such is based on an externally managed reference process, which has its own life cycle in which the process is maintained. A new "V-Modell XT Bund" PR thus causes an update trigger generating a change request in the agency's change management system. In the next iteration, externally caused change requests become improvement requirements. ArSPI provides the agency with information of how the process variant was derived from the reference process (e.g., in the design artefacts, in the *SPLDeltaReport* artefact, etc.) and, thus, helps to determine the changes of the reference process and how these changes affect the own variant (e.g., changed process assets and transitively affected ones). Due to the fine-grained artefact model, affected artefacts can be identified, and respective work packages to adopt changes can be defined. As Figure 4 also shows, the "V-Modell XT Bund" itself is a variant of the "V-Modell XT Reference Model" and, thus, has the same situation of internally and externally triggered evolution.

### 3.2.2 SPI Project Point of View

Having defined the *Project Assignment*, the SPI project is initiated following the life cycle phases (Figure 1, Figure 3, and Table 1). Each phase produces at least one key artefact, which contains and structures the analysis results, the process designs, life cycle support documentation, and releases. Based on a *Vision*, a set of *Changes*, and an *Actual Process*, process engineers start to elaborate the requirements relevant for the actual improvement cycle, and they assemble them in the *Process Requirements*. Based on the requirements, the process designs are created. ArSPI proposes a two-staged design process (reflected by the artefacts *Conceptual Process Design* and *Technical Process Design*) to separate conceptualisation and (technical) realisation. However, as SPI projects can be performed on a small scale, *Conceptual-* and *Technical Process Design* can be integrated into a unified *Process Design* artefact. Finally, the *Process Release* is created, packaged, and shipped to the organisation. Based on the key artefacts, several optional artefacts are created in the SPI project, e.g., plans, assessments, and process-supporting tools.

---

[2] IceScrum web site: http://www.icescrum.org/

**Example.** This example is extracted from an improvement project in Eastern Europe in which a new process was defined supporting project management and development processes (cf. Table 3, study 2). To ground the process in a proven platform, it was an intensively customised variant of the German V-Modell XT. The SPI project (Figure 5) was conducted in 2012/2013, was executed in three iterations, and took about 12 months.

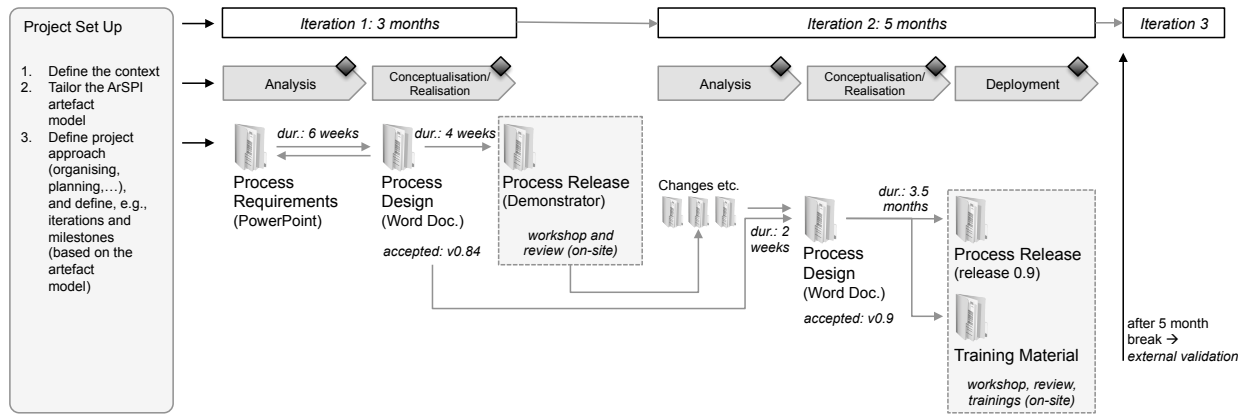

**Figure 5 Example SPI project structure.**

In the following, we exemplarily describe the project set up and the first two iterations. As ArSPI relies on an iterative/incremental approach, the first two iterations give an understanding of the tasks, which are then refined and repeated.

- **SPI Project Set Up:** The set up serves the tailoring of the artefact model for a particular SPI project. Few experience-based questions (e.g., for the context, pre-knowledge, and training strategies) need to be answered to support the determination of the relevant artefacts and their representation. For instance, Figure 5 shows the artefact type *Process Design*, which is a simplified artefact that integrates the *Conceptual-* and *Technical Process Design* artefacts addressing small SPI projects. Finally, the overall project approach is defined, e.g., by defining milestones, and by creating the project-related manuals such as for project- or quality management procedures.
- **Iteration 1:** As shown in Figure 5, the first step consists of conducting the analysis leading to the artefact *Process Requirements*. After the first analyses, the *Process Design* and the *Process Release* (as demonstrator) are created. The figure shows the first iteration to be shortened, as it does not contain a deployment phase. In this project, the selected approach contained a prototyping work package in which the basic requirements should be analysed and prototypically implemented. That is, only a demonstrator should be delivered as *Process Release*, which was then evaluated by the process owners.
- **Iteration 2:** In the second iteration, the *Process Design* is refined in response to the client's evaluation and the change requests. In Figure 5, we show the evolution of the *Process Design*. Furthermore, the second iteration should produce a full *Process Release* for initial deployment and validation purposes. In a 4-day workshop, we prepared the delivery and evaluation, and performed the initial training for early adopters.

This example shows how ArSPI is adopted to a particular SPI project. In the beginning of the project, after determining the project context, the artefact model is adapted to the specific needs, e.g., artefacts are merged or simplified and optional support artefacts that are considered relevant are selected for creation (e.g., *Training Concept*, *Training Plan*, and *Training Material*). Furthermore, based on the life cycle model, a concrete project approach is defined by, e.g., defining milestones, planning iterations, and defining the life cycle phases that are needed for the iterations. As the ArSPI model precisely defines the artefacts, their structure, and the dependencies, process engineers get a set of consistent and complete artefacts for planning and monitoring the SPI projects while preserving the necessary degree of flexibility in the way of working.

### 3.3 Validating ArSPI

We developed an evaluation strategy to validate ArSPI from different angles. A detailed description of the evaluation strategy and the conducted case studies can be taken from [18]. The backbone of the evaluation strategy is a combination of different empirical instruments applied in academic (internal) as well as in industry-hosted (external) settings. The internal validation aims at creating settings to (1) validate ArSPI in controlled environ-ments, (2) to analyse the model's consistency and completeness, and to (3) develop/refine the instruments to be used in the external validation. Furthermore, the internal validation paves the way for independent replication studies. The external validation aims at providing insights into practical settings regarding benefits and shortcomings to prepare dissemination and further investigations. As our approach was developed based on our experiences to systematise the pragmatic approaches in the past [18], the conducted case studies thereby aim at investigating whether and to which extent ArSPI generally supports process engineers in conducting a systematic SPI.

### 3.3.1 Overview of conducted Studies and Outcomes

In Table 3, we give an overview of the overall strategy implemented so far (studies that were recently added to the evaluation from [18] are marked with "new"). We summarise the instruments, context, a brief study description, and outcomes.

**Table 3 Overview of the elements of the validation strategy.**

| No. | Validation | | | | Outcomes | | | |
|---|---|---|---|---|---|---|---|---|
| | Inst | I/E | Ctx | Description | Process | Templates | Tool | Training |
| 1 | Exp | I | U | This quasi-experiment was conducted in a course on software process modelling [22], and aimed at investigating the feasibility of an artefact-based SPI approach in general. The whole experiment is described in [21]. | X | | | |
| 2 | CS | E | I | In this industry-hosted case study, a new process should be developed, which aimed to define management and development procedures. The new process was based on the V-Modell XT. As no case existed in advance, the case study could not be conducted in a comparative manner. However, in order to set up a continuous improvement and management, the process was evaluated using interviews to create reference values for further evaluation. A detailed description can be depicted from [18] and Section 3.2.2. | X | X | | X |
| 3$_{new}$ | CS | I | U | In this investigation, the release 0.9 of ArSPI was analysed for completeness. The overall goal was to define requirements for ArSPI's further improvement based on the experiences gathered so far and using CMMI as external reference. | X | | X | |
| 4$_{new}$ | CS | E | I | In this case study, ArSPI was evaluated from the perspective of the process owner who is responsible for his company's SPM. We combined elements from experimentation and case study research to evaluate ArSPI in a real-world setting, and to provide a controlled environment to gather detailed insights into the execution of the project. For this, two industry partners defined the requirements and acted as clients in the project. A student performed the SPI activities. The objective was to tailor the Scrum process respecting the predefined requirements. We as developers of ArSPI monitored the project and performed a continuous evaluation regarding, e.g., product and process quality. | X | | X | |
| 5 | CS | E | I | This study is a long-term industry case study in which the organisation-wide software process is subject to continuous improvement. In this particular setting, the ArSPI model is applied to SPI projects as well as to the organisation-wide quality- and process management. A description of this setting can be depicted from [18] and Section 3.2.1. | X | X | X | X |

| No. | Validation | | | | Outcomes | | | |
|---|---|---|---|---|---|---|---|---|
| | Inst | I/E | Ctx | Description | Process | Templates | Tool | Training |
| 6 | Int | E | I | During the construction of the ArSPI model, we conducted interviews with external partners from industry and academia, who are experienced in SPI. The interviews aimed at investigating the completeness of the constructed model, and to figure out improvement potential. A description of the interview and its outcomes can be depicted from [18]. | X | | | |
| Legend: <br> Inst – Instrument: Exp = experiment, CS = Case Study, Int = interview; I/E – internal/external validation: I = internal, E = external; <br> Ctx – Context: U = university/academic, I = industry ||||||||

### 3.3.2 Summary of Conclusions

So far, we developed ArSPI in an inductive manner complemented with continuous validation and evaluation activities serving its improvement. From the initially conducted studies, we could extract the following findings:

- Process consumers, e.g., process owners or tool developers, benefit from an artefact-based SPI approach as the artefact-based approach allows for, e.g., a precise definition of process entities for tool support or process enactment [21]. A major finding was that we can rate the success of an SPI project by rating the outcome, i.e. we imply a notion of SPI quality in relation to the quality of the outcome while abstracting from the way of producing the outcome – which is the fundamental principle of artefact orientation.
- Process engineers benefit from an artefact-based SPI approach by being provided with a clearly structured model serving as reference to design/improve processes [18], [21]. For example, in study 2 (Table 3), the evaluation of release 0.9 of the developed process indicated to gaps, which could be directly aligned to change requests; process owners mentioned missing artefacts and 5 missing artefacts could be identified. As figured out in [21], we can rate the quality of SPI projects by rating the outcomes.
- Experts consider ArSPI useful, as, for instance, it helps to structure SPI projects, and to reflect on SPI activities [18]. A major finding was the flexibility of the ArSPI model that allows for tailoring and applying ArSPI in different contexts, e.g., large and small, and short- and long-term SPI projects/programs.

However, the number and character of the conducted case studies limit our initial findings. For instance, so far, completed case studies mainly address stakeholders related to process management, and, thus, project managers and software developers were not in scope as primary study subjects. However, in a complementing study [20], we could find indication to benefits for these stakeholder groups as well.

### 3.3.3 Exemplary Results

In the studies 3 and 4 (Table 3), we aimed at conducting a comparative in-depth analysis of ArSPI compared to previously used approaches. In the following, we provide insight into the industry-hosted study 4 in which we conducted a completely monitored case study. Two industry partners were personally invited to participate in the case study and were asked to rate the ArSPI approach in relation to their experiences.

Figure 6 illustrates the final rating of the experts as an exemplarily evaluation of ArSPI (the ratings are based questionnaires and interviews, values are on an 8-point Likert scale). Expert 1 has experienced 6 medium- to large-sized SPI projects, mainly in the context of public administration. Expert 2 has conducted about 50 SPI and SPI-related projects in different industry contexts.

Figure 6 shows that especially expert 2 rates the approach significantly better than the previously experienced ones. He stated that although there were some limitations by the study's setting, the ArSPI approach worked "better that everything else compared to what happens in practice right now." The evaluation of expert 1 shows a different picture. Expert 1 also rates structuredness, knowledge transfer, and explicit analysis and design procedures "good" (5 to 6). However, based on his experiences, he gave a low rating for the other criteria resulting from "the way the process engineer applied the model in this case study."

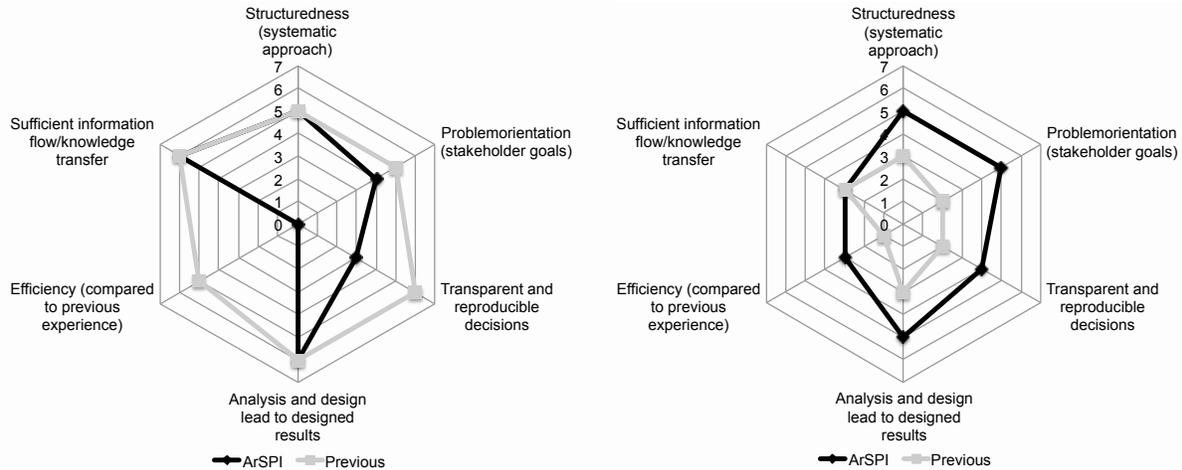

**Figure 6 Comparative expert evaluation form study 4 (expert 1: left, expert 2: right).**

From these evaluations, we draw the conclusion that, on the one hand, ArSPI is considered a useful instrument, which, on the other hand, needs to be extended by more guidance to further improve process engineer support. The observations made in this case study also comply with our arguments for a better education of process engineers [22], and also support the need for a comprehensive set of SPI success factors, which are subject to current SPI-related research. However, as these aspects were not in the focus of the conducted study, the respective findings call for further investigations.

## 4 Conclusion

In this paper, we presented a model for Artefact-based Software Process Improvement & Management (called ArSPI), which emerged from several SPI projects. Due to the complexity of the SPI projects and the context, we experienced the demand for a systematic approach that allows for structuring SPI projects, creating reusable process assets, and for checking process requirements, designs, and realisations for consistency. Therefore, we extracted best practices and SPI artefacts to develop an approach to set up, organise, and perform SPI projects. The approach allows for defining a structure for SPI project outcomes and gives process engineers the freedom to select concrete methods to analyse, design, and implement processes in relation to the particular situation. Our approach thereby provides a reference to organise and manage systematic SPI projects/programs while the underlying artefact-based approach does not restrict the actual SPI endeavour by normative, solution-driven ways of working that might be alien with the organisational culture.

An evaluation strategy combining internal validation in academic and external evaluation in industry context served the determination of strengths and weaknesses. The results from our studies indicate that ArSPI serves as useful reference defining a structured guidance to systematically set up problem-driven SPI projects.

A major threat to the external validity and, thus, to the ability to generalise is that, besides inferring our approach from experiences, we applied the approach in four industry projects, and conducted so far two case studies. A threat inherent to case study research is that the boundaries between the project characteristics and the phenomena are unclear whereby it might remain unknown whether certain effects are directly caused by applying ArSPI or by other unknown side-effects, e.g., a generally more structured way of working. Our investigations lay, however, a first and necessary basis to further investigate the effects of applying ArSPI under realistic conditions. With the instruments and the published material, we also lay the foundation for future replications. These are necessary to get in-depth insights into the benefits and limitations of applying ArSPI in practice and we postulate that those replications need to be conducted independently. This is also necessary to answer the question whether the success of ArSPI results from us applying the concepts. To manifest our results, we therefore need further studies whereas we laid with our contributions at hands the necessary foundation.

## 4.1 Impact and Implications

The ArSPI model provides a blueprint for setting up and organising SPI projects. The model, its documentation, and templates are available online. ArSPI is focused on the artefacts needed by process engineers to analyse process requirements, to design and implement processes, and to ship processes and establish a continuous improvement. Since ArSPI is focused on artefacts, process engineers can directly apply the model to structure SPI activities.

Researchers and practitioners as well get with our contribution already insights into benefits and shortcomings in SPI in general and in artefact-based SPI in particular. As we created an experimental setting in which SPI activities can be analysed, compared, and evaluated, we actively contributed to the dissemination into academia and practice and support to the replications of our studies and to further expand our knowledge on the broad spectrum of SPI knowledge.

## 4.2 Future Work

ArSPI still needs a continuous validation to foster its improvement. Beyond an initial Eclipse Process Framework-based implementation of ArSPI, we are also working on implementations using other frameworks and on the development of further supporting material, e.g. checklists, evaluation questionnaires, etc. Findings from the conducted studies become part of the next iteration of ArSPI, e.g., findings from recent studies 3 and 4 (Section 3.3.3) define improvement requirements. Furthermore, ArSPI is under analysis for integration opportunities with existing standards, e.g., the ISO/IEC 33014 [16]. In addition to the practical dissemination, we also plan to extend the process-engineering lab [21] to systematically analyse and understand findings from practical studies. Those different steps serve the dissemination of our approach and, especially, the continuous, joint evaluation of ArSPI to which we cordially invite researchers and practitioners.

## Acknowledgements

We owe special thanks to Sarah Beecham and Ita Richardson for the fruitful discussion on SPI methods. Furthermore, we thank Jens Calamé for reviewing the article manuscript.

## 5 References

[1] Aysolmaz, B., Demirörs, O.: 'A Detailed Software Process Improvement Methodology: BG-SPI', Proc. European Conf. System & Software Process Improvement and Innovation, Roskilde, Denmark, June 2011, pp. 97–108

[2] Beecham, S.: 'A Requirements-based Software Process Maturity Model'. PhD Thesis, Department of Computer Science, University of Hertfordshire, UK, 2003

[3] Birk, A., Pfahl, D.: 'A Systems Perspective on Software Process Improvement', Proc. Int. Conf. Product Focused Software Process Improvement, Rovaniemi, Finnland, December 2002, pp. 4–18

[4] Boria, J. L.: 'A framework for understanding software process improvement's return on investment', Proc. Portland Int. Conf. on Management and Technology Innovation in Technology Management, Portland, USA, July 1997, pp. 847–851

[5] CMMI Product Team, 'CMMI for Development, Version 1.3' (Software Engineering Institute, Carnegie Mellon University, 2010)

[6] Coleman, G., O'Connor, R.: 'Investigating software process in practice: A grounded theory perspective', Journal of Systems and Software, 2008, 81, (5), pp. 772–784

[7] Deming, W. E. Out of the Crisis. MIT Press, 2000.

[8] Eberlein, A., Jiang, L.: 'Description of a process development methodology', Software Process: Improvement and Practice, 2007, 12, (1), pp. 101–118

[9] Frailey, D. J.: 'Defining a corporate-wide software process', Proc. Int. Conf. on the Software Process, Redondo Beach, USA, October 1991, pp. 113–121

[10] Hardgrave, B. C., Armstrong, D. J.: 'Software process improvement: It's a journey not a destination', Communications of the ACM, 2005, 48, (11), 93–96

[11] Horvat, R. V., Rozman, I., Györkös, J.: 'Managing the Complexity of SPI in Small Companies', Software Process: Improvement and Practice, 2000, 5, (1), pp. 45–54
[12] Humphrey, W. S.: 'Managing the Software Process' (Addison Wesley, 1989)
[13] ISO, 'ISO/IEC 12207:2008: Systems and software engineering – Software life cycle processes' (International Organization for Standardization, 2008)
[14] ISO, 'ISO/IEC 15504-4:2004: Software Process Assessment - Part 4: Guidance on use for process improvement and process capability determination' (International Organization for Standardization, 2004)
[15] ISO, 'ISO/IEC 29110:2011: Systems and Software Life Cycle Profiles and Guidelines for Very Small Entities (VSEs)' (International Organization for Standardization, 2011)
[16] ISO, 'ISO/IEC 33014:2013: Information Technology – Process Assessment – Guide for Process Improvement' (International Organization for Standardization, 2013)
[17] Kuhrmann, M.: 'ArSPI: An Artifact Model for Software Process Improvement and Management' (Technische Universität München, 2013)
[18] Kuhrmann, M.: 'Crafting a Software Process Improvement Approach – A Retrospective Systematization', Journal of Software: Evolution and Process, 2015, 27, (2), pp. 114–145
[19] Kuhrmann, M., Beecham, S.: 'Artifact-based Software Process Improvement and Management: A Method Proposal', Proc. Int. Conf. on Software and Systems Process, Nanjing, China, May 2014, pp. 119–123
[20] Kuhrmann, M., Lange, C., Schnackenburg, A.: 'A Survey on the Application of the V-Modell XT in German Government Agencies', Proc. European Conf. System & Software Process Improvement and Innovation, Roskilde, Denmark, June 2011, pp. 49–60
[21] Kuhrmann, M., Méndez Fernández, D., Knapp, A.: 'A First Investigation About the Perceived Value of Process Engineering and Process Consumption', Proc. Int. Conf. Product Focused Software Process Improvement, Paphos, Cypros, June 2013, pp. 138–152
[22] Kuhrmann, M., Méndez Fernández, D., Münch, J.: 'Teaching Software Process Modeling', Proc. Int. Conf. on Software Engineering, San Francisco, USA, May 2013, pp. 1138–1147
[23] Kuhrmann, M., Méndez Fernández, D., Ternité, T.: 'Realizing Software Process Lines: Insights and Experiences', Proc. Int. Conf. on Software and Systems Process, Nanjing, China, May 2014, pp. 110–119
[24] Kuvaja, P.: 'Bootstrap 3.0 – A Spice Conformant Software Process Assessment Methodology', Software Quality Journal, 1999, 8, (1), pp. 7–19
[25] Mendéz Fernández, D., Penzenstadler, B., Kuhrmann, M., Broy, M.: 'A Meta Model for Artefact-Orientation: Fundamentals and Lessons Learned in Requirements Engineering', Proc. Int. Conf. on Model Driven Engineering Languages and Systems, Oslo, Norway, October 2010, pp. 183–197
[26] Mendéz Fernández, D., Wieringa, R.: 'Improving Requirements Engineering by Artefact Orientation', Proc. Int. Conf. Product Focused Software Process Improvement, Paphos, Cypros, June 2013, pp. 108–122
[27] Münch, J., Armbrust, O., Sotó, M., Kowalczyk, M.: 'Software Process Definition and Management' (Springer, 2012)
[28] Niazi, M., Wilson, D., Zowghi, D.: 'A maturity model for the implementation of SPI', Journal of Systems and Software, 2003, 74, (2), pp. 155–172
[29] Ocampo, A., Münch, J.: 'Rationale Modeling for Software Process Evolution', Journal on Software Process: Improvement and Practice, 2009, 14, (2), pp. 85–105
[30] Rainer, A., Hall, T.: 'An analysis of some 'core studies' of software process improvement', Software Process: Improvement and Practice, 2002, 6, (4), pp. 169–187
[31] Raninen, A., Ahonen, J. J., Sihvonen, H.-M., Savolainen, P., Beecham, S.: 'LAPPI: A light-weight Technique to Practical Process Modeling and Improvement Target Identification', Journal of Software: Evolution and Process, 2012, 25, (9), pp. 915–933
[32] Staples, M., Niazi, M., Jeffery, R., Abrahams, A., Byatt, P., Murphy, R.: 'An Exploratory Study of Why Organizations do not Adopt CMMI', Journal of Systems and Software, 2007, 80, (6), pp. 883–895
[33] Stelzer, D., Mellis, W.: 'Success Factors of Organizational Change in Software Process Improvement', Software Process: Improvement and Practice, 1998, 4, (4), pp. 227–250

[34] Viana, D., Conte, T., Vilela, D., De Souza, C. R. B., Santos, G., Prikladnicki, R.: 'The influence of human aspects on software process improvement: Qualitative research findings and comparison to previous studies', Proc. Int. Conf. on Evaluation & Assessment in Software Engineering, Ciudad Real, Spain, May 2012, pp. 121–125